\documentstyle[aps]{revtex}
\tighten
\begin{document}


\preprint{CITA-2001-36,gr-qc/0108048}

\title{Doubly covariant action principle of singular hypersurfaces in
general relativity and scalar-tensor theories}
\author{Shinji Mukohyama}
\address{
Department of Physics, Harvard University\\
Cambridge, MA 02138, USA\\
Canadian Institute for Theoretical Astrophysics, 
University of Toronto\\
Toronto, ON M5S3H8, Canada\\
Department of Physics and Astronomy, 
University of Victoria\\ 
Victoria, BC V8W3P6, Canada
}
\date{\today}

\maketitle


\begin{abstract} 

 An action principle of singular hypersurfaces in general relativity and
 scalar-tensor type theories of gravity in the Einstein frame is
 presented without assuming any symmetry. The action principle is
 manifestly doubly covariant in the sense that coordinate systems on and
 off a hypersurface are disentangled and can be independently
 specified. It is shown that, including variation of the metric, the
 position of the hypersurface and matter fields, the variational
 principle gives the correct set of equations of motion: the Einstein
 equation off the hypersurface, Israel's junction condition in a doubly
 covariant form and equations of motion of matter fields including the
 scalar fields. The position of the hypersurface measured from one side
 of the hypersurface and that measured from another side can be
 independently variated as required by the double covariance.

\end{abstract}



\section{Introduction}


Spatially extended objects such as branes, membranes, shells and walls 
have been playing many important roles in recent progress in theoretical
physics including string theory~\cite{Polchinski}, particle
phenomenology~\cite{ADD,AAD,RS1}, theory of black
holes~\cite{TPM,PI,SV}, cosmology~\cite{KT,Linde} and so on. Hence, it
is important to investigate dynamics of such extended objects. In
particular, the so called brane world scenario is based on the idea that
our four-dimensional universe may be a world volume of a brane in a
higher dimensional spacetime~\cite{ADD,AAD,RS1,RS2}. Thus, in the brane
world scenario the dynamics of the brane is the dynamics of our universe
itself and is of the most physical importance.


It is well-known and is the most commonly adopted picture that the
dynamics of extended objects is elegantly described as geometrical
imbedding of world-volume surfaces into spacetime in a certain 
limit. In particular, in the case of codimension $1$, or when the
world-volume surface is a hypersurface, the geometrical description
becomes simpler than other cases with higher codimension. Actually, in
general relativity or other theories of gravity in the Einstein frame,
the classical dynamics of a hypersurface is perfectly described by
Israel's junction condition~\cite{Israel}.


One of the main advantages of the junction condition is that it is
manifestly doubly covariant in the sense that coordinate systems on and
off a hypersurface are disentangled and can be independently
specified. More precisely, there are three independent coordinate
systems: that on the hypersurface, those in two regions separated by the
hypersurface. In the brane world point of view, the double covariance is
important since it allows us to separate the coordinate system in our
world from that in the higher dimensional spacetime.


Once the classical dynamics is understood, one would usually like to
understand quantum mechanical
dynamics~\cite{BKKT,Berezin,Hajicek,HKK,NOT,DK,HB,HL,FLW,Gladush}. For
this purpose, we would like to obtain the action principle for the
system including a hypersurface.


The easiest way to obtain the action may be to adopt the Gaussian normal
coordinate system based on the hypersurface and to consider the
Einstein-Hilbert action with a delta function source. The action
obtained in this way gives the correct set of equations in the
coordinate system, provided that the position of the hypersurface and
coordinates in a neighborhood of the hypersurface are fixed by the
Gaussian normal coordinate condition. However, in this method we lose
the double covariance: coordinates on the hypersurface is a part of
coordinates off the hypersurface so that the coordinates satisfy the 
Gaussian normal coordinate condition. The loss of the double covariance
is regrettable.


Actually, as far as the author knows, a doubly covariant action
principle has not yet been obtained in the literature. One of the main
difficulties seems due to the fact that the spacetime metric on one side
of the hypersurface and that on another side are independent variables in
the variational principle. Hence, a question arises: How can we ensure
the regularity of the intrinsic geometry of the hypersurface without 
entangling the coordinate systems on and off the hypersurface? This
question will be answered in this paper as a manifestly covariant action
principle will be presented.


Another difficulty is due to the fact that the double covariance
requires inclusion of the position of the hypersurface as a dynamical 
variable in the action principle. Actually, in the doubly covariant
formulation of the junction condition, it is easy to see that variables
specifying the position are not invariant under coordinate
transformation and should not be fixed~\cite{Mukohyama2000c}. More about
why we need to include the position of the hypersurface will be
explained from the brane world point of view in
Sec.~\ref{sec:summary}. Here, we mention that, since coordinate systems
in two sides of the hypersurface are independent, the position of the
hypersurface measured from one side and that measured from another side
should be independently variated in the variational principle.


It may be worth while reviewing the present status in the literature
regarding the second difficulty. However, the author knows only a few
papers referring this point. Here, we only quote a sentence from one
of them: `{\it The variational equations that arose from the unreduced
Hamiltonian action were not strictly consistent in a distributional
sense, but we were able to localize the ambiguity into the single
equation that arises by varying the action with respect to the shell
position}~\cite{FLW}`. One might think that another paper~\cite{Gladush}
had obtained the correct set of equations, but in that paper the
position of the hypersurface measured from one side and that measured
from another side can not be variated independently. Actually, if we
would simply variate the position of the hypersurface measured from one
side and that measured from another side independently, then the
variational principle presented in ref.~\cite{Gladush} would give wrong
equations. Moreover, in both of these papers, the hypersurface
represents only a dust shell and the coordinate systems on and off the
hypersurface are not independent. One of them~\cite{FLW} assumes
spherical symmetry, too.


The purpose of this paper is to provide a manifestly doubly covariant
action principle of singular hypersurfaces in general relativity and 
scalar-tensor type theories of gravity in the Einstein frame without
assuming any symmetry. Besides the scalar fields included in the
scalar-tensor type theories, any kind of matter Lagrangian density on
the hypersurface, which may depend also on the pullback of the scalar
fields, can be included. It is shown that, including variation of the
metric, the position of the hypersurface and matter fields, the
variational principle gives the correct set of equations of motion: the
Einstein equation off the hypersurface, Israel's junction condition in a
doubly covariant form and equations of motion of matter fields including
the scalar fields. As required by the double covariance, the position of
the hypersurface measured from one side of the hypersurface and that
measured from another side can be independently variated.


This paper is organized as follows. In Sec.~\ref{sec:action} a doubly
covariant action of a singular hypersurface is derived from the standard
Einstein-Hilbert action. In Sec.~\ref{sec:variation} the variation of
the action is calculated for the variations of the metric and the
position of the hypersurface, and the corresponding equations are
obtained. In Sec.~\ref{sec:examples} the variation of the action
corresponding to the variations of scalar fields are
evaluated. Sec.~\ref{sec:summary} is devoted to a summary of this paper
and some discussions.


\section{Action of singular hypersurface}
\label{sec:action}

Let us consider a $D$-dimensional spacetime $({\cal M},g_{MN})$ and 
a timelike or spacelike hypersurface $\Sigma$ which separates ${\cal M}$
into two regions, ${\cal M}_+$ and ${\cal M}_-$. Since we would 
like to consider $\Sigma$ as a physical object (eg. the world-volume of
a brane or the world-volume of a bubble wall in a first-order phase
transition) or a physical event (eg. instantaneous global phase
transition~\cite{FMM}), we assume that the $(D-1)$-dimensional intrinsic geometry
on $\Sigma$ is regular. On the other hand, the $D$-dimensional geometry
is not necessarily regular on $\Sigma$.

In the following arguments we shall estimate the action for the system
including the singular hypersurface $\Sigma$. We assume that the system
is described by the action~\footnote{For simplicity we do not consider
the boundary of ${\cal M}$, but it is easy to take it into account by
imposing suitable boundary conditions and introducing boundary terms
appropriate for the boundary condition.} 
%
\begin{equation}
 I_{tot} = I_{EH}+I_{matter},
  \label{eqn:total-action}
\end{equation}
where $I_{EH}$ is the Einstein-Hilbert action with a cosmological
constant
%
\begin{equation}
 I_{EH} = \frac{1}{2\kappa^2}\int_{\cal M}d^Dx\sqrt{|g|}(R-2\Lambda),
\end{equation}
and $I_{matter}$ is the matter action of the form
%
\begin{equation}
 I_{matter} = \int_{{\cal M}_+}d^Dx_+{\cal L}_+
  +\int_{{\cal M}_-}d^Dx_-{\cal L}_-
  +\int_{\Sigma}d^{D-1}y{\cal L}_0.\label{eqn:matter-action}
\end{equation}
Here, $\{x^M_{\pm}\}$ are $D$-dimensional coordinate systems in ${\cal
M}_{\pm}$, respectively, and $\{y^{\mu}\}$ is a $(D-1)$-dimensional
coordinate system in $\Sigma$. The three coordinate systems can be
independent from each other.

In order to evaluate the gravitational part of the action, we first
regularize the $D$-dimensional geometry in a neighborhood of 
$\Sigma$ by introducing the finite thickness $\delta$ of the object
corresponding to $\Sigma$. Of course, in the final step below, we shall
take the limit $\delta\to+0$, where the hypersurface becomes singular
again. Namely, we consider the prescription
%
\begin{eqnarray}
 I_{EH} & = & \lim_{\delta\to+0}
  (I_+^{\delta}+I_0^{\delta}+I_-^{\delta}),\nonumber\\
 I_0^{\delta} & = &
  \frac{1}{2\kappa^2}\left[\int_{{\cal M}_0^{\delta}} 
  d^Dx\sqrt{|g|}(R-2\Lambda)
  + 2\epsilon \int_{{\cal B}_+^{\delta}}d^{D-1}y\sqrt{|q|}K
  - 2\epsilon \int_{{\cal B}_-^{\delta}}d^{D-1}y\sqrt{|q|}K\right],
  \nonumber\\
 I_{\pm}^{\delta} & = &
  \frac{1}{2\kappa^2}\left[\int_{{\cal M}_{\pm}^{\delta}} 
  d^Dx\sqrt{|g|}(R-2\Lambda)
  \mp 2 \epsilon\int_{{\cal B}_{\pm}^{\delta}}d^{D-1}y 
  \sqrt{|q|}K\right], 
  \label{eqn:regularization}
\end{eqnarray}
where ${\cal M}_0^{\delta}$ is a spacetime neighborhood of $\Sigma$
representing the regularized object, ${\cal M}_{\pm}^{\delta}$ are
the two regions separated by ${\cal M}_0^{\delta}$ so that 
%
\begin{equation}
 {\cal M}_0^{\delta}\supset \Sigma,\ 
 {\cal M}_{\pm}^{\delta}\subset {\cal M}_{\pm},\ 
  \lim_{\delta\to+0}{\cal M}_{\pm}^{\delta} = {\cal M}_{\pm},
\end{equation}
and ${\cal B}_{\pm}^{\delta}$ is the boundary between 
${\cal M}_0^{\delta}$ and ${\cal M}_{\pm}^{\delta}$, respectively. 
Note that surface terms have been included in $I_{0,\pm}^{\delta}$ for
later convenience but that these exactly cancel each other on common
boundaries ${\cal B}_{\pm}^{\delta}$. Each surface term is defined as an
integral over the $(D-1)$-dimensional intrinsic coordinates $y^{\mu}$ on
${\cal B}_{\pm}^{\delta}$, $q$ is the determinant of the
induced metric, $K$ is the trace of the extrinsic curvature
associated with the unit normal $n^M$ directed from 
${\cal M}_0^{\delta}$ to ${\cal M}_+^{\delta}$ or from 
${\cal M}_-^{\delta}$ to ${\cal M}_0^{\delta}$, and
$\epsilon=g_{MN}n^Mn^N=\pm 1$.

Next, in order to estimate $I_0^{\delta}$, we foliate 
${\cal M}_0^{\delta}$ by such a one-parameter family of hypersurfaces
$\Sigma_{\tau}$ that $\Sigma_{0}$ coincides with $\Sigma$ and that
$\Sigma_{\pm 1}$ coincides with the boundary ${\cal B}_{\pm}^{\delta}$,
respectively. Hence, we can decompose the $D$-dimensional Ricci scalar
$R$ as 
%
\begin{equation}
 R = R^{(D-1)} + \epsilon K^2 - \epsilon K^{\mu\nu}K_{\mu\nu} 
  - 2\epsilon (Kn^M-n^M_{;N}n^N)_{;M},
\end{equation}
where $R^{(D-1)}$ is the Ricci scalar of the $(D-1)$-dimensional
induced metric on $\Sigma_{\tau}$, the semicolon represents the
covariant derivative compatible with $g_{MN}$, $n^M$ is the unit normal
to $\Sigma_{\tau}$ directed towards ${\cal B}_+^{\delta}$, 
$\epsilon=g_{MN}n^Mn^N=\pm 1$, $K_{\mu\nu}$ is the extrinsic curvature
associated with $n^M$, the indices $\{\mu,\nu\}$ are raised by the
inverse of the induced metric, and $K=K^{\mu}_{\nu}$. By integrating
over ${\cal M}_0^{\delta}$ and taking the limit $\delta\to+0$, we obtain
%
\begin{equation}
 I_0^{\delta} = \frac{1}{2\kappa^2}
  \int_{{\cal M}_0^{\delta}}d^Dx\sqrt{|g|}
  \left( R^{(D-1)} + \epsilon K^2 - \epsilon K^{\mu\nu}K_{\mu\nu}
   -2\Lambda\right)
 \to 0\ (\delta\to+0).
\end{equation}
Here, we have used the assumption that the intrinsic geometry on
$\Sigma$ is regular even in the limit $\delta\to+0$. We have also
assumed that the extrinsic curvature remains finite.

Therefore, we obtain the following form of the Einstein-Hilbert action
for the system including the singular hypersurface $\Sigma$. 
%
\begin{eqnarray}
 I_{EH} & = & \frac{1}{2\kappa^2}\left[
  \int_{{\cal M}_+}d^Dx_+\sqrt{|g_+|}(R_+-2\Lambda_+)
   + \int_{{\cal M}_-}d^Dx_-\sqrt{|g_-|}(R_--2\Lambda_-)\right.
   \nonumber\\
 & & \left.-2\epsilon\int_{\Sigma}d^{D-1}y
       \left(\sqrt{|q_+|}K_+ - \sqrt{|q_-|}K_-\right)
       +\epsilon\int_{\Sigma}d^{D-1}y
       \lambda^{\mu\nu}(q_{+\mu\nu}-q_{-\mu\nu})
       \right],
       \label{eqn:EH-action}
\end{eqnarray}
where $q_{\pm\mu\nu}$ is the induced metric, $q_{\pm}$ is the
determinant of $q_{\pm\mu\nu}$, $K_{\pm}=q_{\pm}^{\mu\nu}K_{\pm\mu\nu}$
is the trace of the extrinsic curvature $K_{\pm\mu\nu}$ and
$q_{\pm}^{\mu\nu}$ is the inverse of $q_{\pm\mu\nu}$. 
In the expression (\ref{eqn:EH-action}) we have distinguished
geometrical quantities in ${\cal M}_+$ and ${\cal M}_-$ by introducing
the subscript $\pm$, and have allowed the cosmological constant to have
different values in these two regions. We have introduced the Lagrange
multiplier field $\lambda^{\mu\nu}(y)$ to ensure the regularity of the
intrinsic geometry of $\Sigma$. When we regularized the system and
decomposed $I_{EH}$ into $I_0^{\delta}$ and $I_{\pm}^{\delta}$ as in
(\ref{eqn:regularization}), we implicitly assumed that the induced 
metric and the extrinsic curvature are continuous across the boundaries
${\cal B}_{\pm}^{\delta}$. After taking the limit $\delta\to+0$, the
extrinsic curvature remains finite but can be discontinuous across
$\Sigma$. On the other hand, the induced metric should remain continuous
across $\Sigma$ even after taking the limit $\delta\to+0$ because of the
finiteness of the extrinsic curvature. Provided that the hypersurface
$\Sigma$ is specified as the boundary of ${\cal M}_{\pm}$ by the
parametric equation 
%
\begin{equation}
 x^M_{\pm}=Z_{\pm}^M(y^{\mu}),\label{eqn:sigma-eq}
\end{equation}
the induced metric and the extrinsic curvature are given by
%
\begin{eqnarray}
 q_{\pm\mu\nu}(y) & = & \left.e^M_{\pm\mu}(y)e^N_{\pm\nu}(y)
		      g_{\pm MN}\right|_{x_{\pm}=Z_{\pm}(y)},
  \nonumber\\
 K_{\pm\mu\nu}(y) & = & \frac{1}{2}\left.e^M_{\pm\mu}(y)e^N_{\pm\nu}(y)
	{\cal L}_{n_{\pm}}g_{\pm MN}\right|_{x_{\pm}=Z_{\pm}(y)},
\end{eqnarray}
where $e^M_{\pm\nu}$ are vectors tangent to $\Sigma$ defined by 
%
\begin{equation}
 e^M_{\pm\mu}(y) = \frac{\partial Z_{\pm}^M}{\partial y^{\mu}},
\end{equation}
and $n_{\pm}^M$ is the unit normal to $\Sigma$ directed from 
${\cal M}_-$ to ${\cal M}_+$. To be precise, $n^M_+$ is the
inward-directed unit normal to $\Sigma$ as the boundary of ${\cal M}_+$
and $n^M_-$ is the outward-directed unit normal to $\Sigma$ as the
boundary of ${\cal M}_-$.

Finally, the total action of the system is given by
(\ref{eqn:total-action}), where $I_{EH}$ and $I_{matter}$ are given by
(\ref{eqn:EH-action}) and (\ref{eqn:matter-action}), respectively.


\section{Variational principle}
\label{sec:variation}

In this section we derive equations of motion from the variational
principle based on the action obtained in the previous section. Namely,
we shall extremize the action $I_{tot}$ with respect to the variation
%
\begin{eqnarray}
 g_{\pm MN}(x) & \to & g_{\pm MN}(x) + \delta g_{\pm MN}(x), 
  \nonumber\\
 Z_{\pm}^M(y) & \to & Z_{\pm}^M(y) + \delta Z_{\pm}^M(y). 
\end{eqnarray}
In the following we omit the subscript $\pm$ unless there is possibility
of confusion.

First, it is easy to show that the integrand of the volume term in
$I_{EH}$ changes as follows. 
%
\begin{equation}
 \sqrt{|g|}(R-2\Lambda) \to 
  \sqrt{|g|}\left[(R-2\Lambda)
		   -(G^{MN}+\Lambda g^{MN})
		   \delta g_{MN} 
	     +(\delta g^{MN}_{\quad\ ;N}-\delta g^{;M})_{;M}
	     + O(\delta^2)\right],
 \label{eqn:delta-R}
\end{equation}
where the semicolon represents the covariant derivative compatible with
the background metric $g_{MN}$ (not with the perturbed metric 
$g_{MN}+\delta g_{MN}$), the indices $M,N,\cdots$ are lowered and raised
by the background metric $g_{MN}$ and its inverse $g^{MN}$, and 
$\delta g$ is defined by $\delta g=\delta g^M_M$. Hence, 
%
\begin{eqnarray}
 \delta \int_{{\cal M}_{\pm}}d^Dx_{\pm}\sqrt{|g|}
  (R-2\Lambda) & = &
  -\int_{{\cal M}_{\pm}}d^Dx_{\pm}\sqrt{|g|}
  (G^{MN}+\Lambda g^{MN})\delta g_{MN}
  \nonumber\\
  & & \mp \epsilon\int_{\Sigma}d^{D-1}y \sqrt{|q|}n^M
   \left.(\delta g^N_{M;N}-\delta g_{;M})
  \right|_{x_{\pm}=Z_{\pm}}\nonumber\\
  & & \mp \epsilon\int_{\Sigma}d^{D-1}y \sqrt{|q|}n_M
   \delta Z_{\pm}^M
   \left.(R-2\Lambda)\right|_{x_{\pm}=Z_{\pm}}
   \label{eqn:delta-bulk-action}
\end{eqnarray}
The second term in the right hand side came from the total derivative in
(\ref{eqn:delta-R}) and the last term is due to the change of the region
to be integrated over.

Next, let us consider the surface term in $I_{EH}$. As shown in
ref.~\cite{Mukohyama2000c} the variations of the induced metric and the
extrinsic curvature are given by
%
\begin{eqnarray}
 \delta q_{\mu\nu} & = &
  e^M_{\mu}e^N_{\nu}
  (\delta g_{MN}+\delta Z_{M;N}+\delta Z_{N;M}),\nonumber\\
 \delta K_{\mu\nu} & = &
        \frac{\epsilon}{2}n^Mn^N
        (\delta g_{MN}+2\delta Z_{M;N})K_{\mu\nu}        
        \nonumber\\
 & &    - \frac{1}{2}n^Le^M_{\mu}e^N_{\nu}
        \left[ 2\delta\Gamma_{LMN}
        + \delta Z_{L;MN} + \delta Z_{L;NM}
        + (R_{L'MLN}+R_{L'NLM})\delta Z^{L'}\right],
	\label{eqn:delta-q-K}
\end{eqnarray}
where the right hand side is evaluated at $x_{\pm}^M=Z_{\pm}^M(y)$ and 
%
\begin{equation}
 \delta\Gamma_{LMN} = \frac{1}{2}
        (\delta g_{LM;N}+\delta g_{LN;M}-\delta g_{MN;L}). 
\end{equation}
In order to make the covariant derivatives of $\delta Z^M$ well-defined,
we have to extend $\delta Z^M$ off $\Sigma$. The expressions
(\ref{eqn:delta-q-K}) are independent of the method of the
extension. For details, see ref.~\cite{Mukohyama2000c}. Hence, 
%
\begin{equation}
 \sqrt{|q|}K \to \sqrt{|q|}K+\delta(\sqrt{|q|}K)+O(\delta^2),
\end{equation}
where
%
\begin{eqnarray}
 \delta(\sqrt{|q|}K)/\sqrt{|q|} & = & 
  - \left(K^{\mu\nu}-\frac{1}{2}Kq^{\mu\nu}\right)\delta q_{\mu\nu}
 + \frac{\epsilon}{2} n^Mn^N(\delta g_{MN}+2\delta Z_{M;N})K
 \nonumber\\
 & & - n^Lq^{\mu\nu}e_{\mu}^Me_{\nu}^N 
  (\delta\Gamma_{LMN} + \delta Z_{L;MN}+R_{L'MLN}\delta Z^{L'}). 
\end{eqnarray}
Combining this with the second term in (\ref{eqn:delta-bulk-action}), we
obtain 
%
\begin{eqnarray}
 n^M(\delta g^N_{M;N}-\delta g_{;M})
  + 2\delta(\sqrt{|q|}K)/\sqrt{|q|} & = & 
  - (K^{\mu\nu}-Kq^{\mu\nu})\delta q_{\mu\nu}
  -2 n^MR_{MN}\delta Z^N
  \nonumber\\
 & & -\frac{1}{\sqrt{|q|}}\left[\sqrt{|q|}q^{\mu\nu}
	n^Me_{\nu}^N(\delta g_{MN}+2\delta Z_{M;N})\right]_{,L}e_{\mu}^L
\end{eqnarray}
where we have used the equations
%
\begin{eqnarray}
 [e_{\mu},e_{\nu}]^M & = & 0, \nonumber\\
 e_{\mu}^M(e_{\nu}^Nn_N)_{;M} & = & e_{\mu}^M(n^Nn_N)_{;M} = 0, 
  \nonumber\\
 q^{\mu\nu}e_{\mu}^Me_{\nu}^N + \epsilon n^Mn^N & = & g^{MN}. 
\end{eqnarray}
Thus, the variation of the Einstein-Hilbert action $I_{EH}$ is given by
%
\begin{eqnarray}
 2\kappa^2\delta I_{EH} & = & 
  -\int_{{\cal M}_+}d^Dx_+\sqrt{|g_+|}(G_+^{MN}+\Lambda_+g_+^{MN})
   \delta g_{+MN}
   -\int_{{\cal M}_-}d^Dx_-\sqrt{|g_-|}(G_-^{MN}+\Lambda_-g_-^{MN})
   \delta g_{-MN}\nonumber\\
  & & +\epsilon\int_{\Sigma}d^{D-1}y\left\{
	\left[\sqrt{|q_+|}(K_+^{\mu\nu}-K_+q_+^{\mu\nu})
	 +\lambda^{\mu\nu}\right]\delta q_{+\mu\nu}
	-\left[\sqrt{|q_-|}(K_-^{\mu\nu}-K_-q_-^{\mu\nu})
	  +\lambda^{\mu\nu}\right]\delta q_{-\mu\nu}
			   \right.\nonumber\\
 & & \qquad
      +2n_+^M\left.(G_{+MN}+\Lambda_+g_{+MN})\right|_{x_+=Z_+}
      \delta Z_+^N
      -2n_-^M\left.(G_{-MN}+\Lambda_-g_{-MN})\right|_{x_-=Z_-}
      \delta Z_-^N \nonumber\\
 & & \left.\qquad
      +(q_{+\mu\nu}-q_{-\mu\nu})\delta\lambda^{\mu\nu}\right\},
\end{eqnarray}

Now let us consider the variation of $I_{matter}$. 
%
\begin{eqnarray}
 2\delta I_{matter} & = & 
  \int_{{\cal M}_+}d^Dx_+\sqrt{|g_+|}T_+^{MN}\delta g_{+MN}
  + \int_{{\cal M}_-}d^Dx_-\sqrt{|g_-|}T_-^{MN}\delta g_{-MN}
  \nonumber\\ 
  & & +\epsilon\int_{\Sigma}d^{D-1}y\left[
	\sqrt{|q|}S^{\mu\nu}\delta q_{\mu\nu}
	+2F_{+M}\delta Z_+^M + 2F_{-M}\delta Z_-^M\right],
\end{eqnarray}
where $q_{\mu\nu}$ is either $q_{+\mu\nu}$ or $q_{-\mu\nu}$, and
%
\begin{eqnarray}
 \sqrt{|g_{\pm}|} T_{\pm}^{MN}(x_{\pm}) & = & 
  2\frac{\delta}{\delta g_{\pm MN}(x_{\pm})}
  \int_{{\cal M}_{\pm}}d^Dx'_{\pm}{\cal L}_{\pm}, \nonumber\\
 \sqrt{|q|} S^{\mu\nu}(y) & = & 
  \left.2\epsilon\frac{\delta}{\delta q_{\mu\nu}(y)}
  \int_{\Sigma}d^{D-1}y'{\cal L}_0\right|_{\delta Z^M=0}, \nonumber\\
 \sqrt{|q|} F_{\pm M}(y) & = & \mp n_{\pm M}
  \left.{\cal L}_{\pm}\right|_{x_{\pm}=Z_{\pm}(y)}
  +\epsilon\left.\frac{\delta}{\delta Z_{\pm}^M(y)}
  \int_{\Sigma}d^{D-1}y'{\cal L}_0\right|_{\delta q_{\mu\nu}=0}.
  \label{eqn:T-F-S}
\end{eqnarray}

Therefore, $\delta I_{tot}=0$ is equivalent to the following set of
equations. 
%
\begin{eqnarray}
 G_{\pm}^{MN}+\Lambda_{\pm}g_{\pm}^{MN} & = & \kappa^2T_{\pm}^{MN},
  \nonumber\\
 q_{+\mu\nu} - q_{-\mu\nu} & = & 0, \nonumber\\
 K_+^{\mu\nu}-K_-^{\mu\nu} & = & 
  - \kappa^2\left(S^{\mu\nu} - \frac{1}{D-2}Sq^{\mu\nu}\right), 
  \label{eqn:EOM} 
\end{eqnarray}
%
\begin{equation}
 F_{\pm N} = \mp n_{\pm}^M\left.T_{\pm MN}\right|_{x_{\pm}=Z_{\pm}},
  \label{eqn:consistency}
\end{equation}
and 
%
\begin{equation}
 \lambda^{\mu\nu} = 
  -\sqrt{|q|}(K_{\mp}^{\mu\nu}-K_{\mp}q^{\mu\nu}). 
\end{equation}
In the right hand side of the last equation, the subscript $-$ (or $+$)
should be taken when ${\cal L}_{0}$ is written in terms of $q_{+\mu\nu}$
(or $q_{-\mu\nu}$, respectively).

Note that the equations (\ref{eqn:EOM}) are the Einstein equation and
Israel's junction condition~\cite{Israel}. The last equation is just to
determine the Lagrange multiplier field $\lambda^{\mu\nu}$. Although the
equation (\ref{eqn:consistency}) looks like a new independent equation,
it will be shown in the next section for simple examples that the
equation is compatible with equations of motion of matter
fields. Therefore, the action principle gives the correct set of
equations: the Einstein equation, Israel's junction condition and
equations of motion of matter fields.


\section{Simple examples}
\label{sec:examples}

In this section we show that for simple examples, the equation
(\ref{eqn:consistency}) is compatible with equations of motion of matter
fields. The first trivial example is the case in which all matter fields
are confined on the hypersurface $\Sigma$. This case includes a shell
with an arbitrary equation of state in a vacuum and the brane world
scenario in a purely gravitational bulk with a bulk cosmological
constant and arbitrary matter fields on the brane. In this case, the
consistency condition (\ref{eqn:consistency}) is trivially satisfied
since ${\cal L}_{\pm}=0$ and ${\cal L}_0$ does not change when
$Z_{\pm}^M$ is changed with $q_{\mu\nu}$ fixed.

As the second example, let us consider a simple case in which there is
only a scalar field other than those matter fields confined on the
hypersurface $\Sigma$. Namely, let us consider the following Lagrangian
densities. 
%
\begin{eqnarray}
 {\cal L}_{\pm} & = & -\sqrt{|g|}\left[
	\frac{1}{2}g^{MN}
  \partial_M\Phi\partial_N\Phi
  + V_{\pm}(\Phi)\right], \nonumber\\
 {\cal L}_0 & = & \bar{\cal L}_0(\phi_+) +\lambda_{\phi}(\phi_+-\phi_-), 
\end{eqnarray}
where $\bar{\cal L}_0$ is the Lagrangian density for matter confined on
$\Sigma$, and $\phi_{\pm}$ is the pullback of $\Phi$ on $\Sigma$ defined
by 
%
\begin{equation}
 \phi_{\pm}(y) = \left.\Phi\right|_{x_{\pm}=Z_{\pm}(y)}. 
\end{equation}
The matter Lagrangian density $\bar{\cal L}_0$ on $\Sigma$ can depend on
$\phi_+$ as well. Note that the Lagrange multiplier field
$\lambda_{\phi}(y)$ is necessary in order that the scalar field should
have single value on $\Sigma$ and that $\Sigma$ should be regular. 
For this example we can easily calculate $T_{\pm}^{MN}$, $S^{\mu\nu}$
and $F_{\pm M}$ as follows. 
%
\begin{eqnarray}
 T_{\pm}^{MN} & = & 
  \partial_M\Phi\partial_N\Phi - g_{MN}
  \left[\frac{1}{2}g^{M'N'}\partial_{M'}\Phi
   \partial_{N'}\Phi + V(\Phi)\right],\nonumber\\
 S^{\mu\nu} & = & 
  \frac{2\epsilon}{\sqrt{|q|}}\left.\frac{\delta}{\delta q_{\mu\nu}(y)} 
  \int_{\Sigma}d^{D-1}y'\bar{\cal L}_0\right|_{\delta \phi_+=0}.
\end{eqnarray}
and 
%
\begin{eqnarray}
 F_{+M} & = & 
  \sqrt{|q|}\left[\frac{1}{2}g^{M'N'}\partial_{M'}\Phi
   \partial_{N'}\Phi + V_{\pm}(\Phi)\right]n_M
  + \epsilon(\partial_{\phi_+}\bar{\cal L}_0
  +\lambda_{\phi})\partial_M\Phi, \label{eqn:F+for-scalar}\\
 F_{-M} & = &
  -\sqrt{|q|}\left[\frac{1}{2}g^{M'N'}\partial_{M'}\Phi
	      \partial_{N'}\Phi 
    + V_{\pm}(\Phi)\right]n_M
  - \epsilon\lambda_{\phi}\partial_M\Phi, 
  \label{eqn:F-for-scalar}
\end{eqnarray}
where the right hand sides of (\ref{eqn:F+for-scalar}) and
(\ref{eqn:F-for-scalar}) are evaluated at
$x_{\pm}^M=Z_{\pm}^M(y)$, respectively. Hence, by using 
%
\begin{equation}
 \lambda_{\phi} = -\partial_{\phi_+}\bar{\cal L}_0 
  -\left.\epsilon\sqrt{|q|} n_+^M\partial_M\Phi\right|_{x_+=Z_+} 
  = -\left.\epsilon\sqrt{|q|} n_-^M\partial_M\Phi\right|_{x_-=Z_-},
\end{equation}
which is a part of equations of motion, it is confirmed that
(\ref{eqn:consistency}) is satisfied. Thus, the consistency condition 
(\ref{eqn:consistency}) is actually compatible with equations of motion
of the scalar field.

It is easy to extend the above analysis to an arbitrary number of scalar
fields.


\section{Summary and Discussion}
\label{sec:summary}


We have presented an action principle of singular hypersurfaces in
general relativity in any dimension without assuming any symmetry. Since
an arbitrary number of scalar fields can be consistently included as
shown in Sec.~\ref{sec:examples}, the action principle is applicable to
a wide class of scalar-tensor type theories of gravity in the Einstein
frame. Besides the scalar fields, any kind of matter Lagrangian density
on the hypersurface, which may depend also on the pullback of the scalar
fields, can be included. The action principle is manifestly doubly
covariant in the sense that coordinate systems on and off a hypersurface
are disentangled and can be independently specified. More precisely,
there are three independent coordinate systems: that on the
hypersurface, those in two regions separated by the hypersurface. We
have shown that, including variation of the metric, the position of the
hypersurface and matter fields, the variational principle gives the
correct set of equations of motion: the Einstein equation off the
hypersurface, Israel's junction condition in a doubly covariant form and
equations of motion of matter fields including the scalar fields. It is
worth while mentioning that the position of the hypersurface measured
from one side of the hypersurface and that measured from another side
can be independently variated as required by the double covariance.


Now let us discuss about application of the doubly covariant action
principle to the brane world scenario.

In refs.~\cite{CGS,FTW,BDEL,Mukohyama2000a,Kraus,Ida,MSM} it was shown
that the standard cosmology can be realized in the Randall-Sundrum brane
world scenario in low energy as far as a spatially homogeneous and
isotropic brane is concerned. After that, many authors investigated
cosmological perturbations in the brane-world 
scenario~\cite{Mukohyama2000b,Mukohyama2000c,Mukohyama2001,KIS,Kodama,Maartens,Langlois,BDBL,Koyama-Soda,LMW,LMSW,BMW}.

In particular, four independent equations for scalar perturbations on
the brane in the plane symmetric ($K=0$) background were derived
recently by the author in ref.~\cite{Mukohyama2001}. The number of
independent equations is the same as in the standard cosmology, and it
was shown that in low energy these sets of equations differ only by the
non-local effects due to gravitational waves in the bulk.

In the derivation of the four equations in ref.~\cite{Mukohyama2001} the
author took advantages of the doubly gauge invariant formalism developed
in refs.~\cite{Mukohyama2000b,Mukohyama2000c}. It was essential that the
formalism includes perturbation of the position of a brane as a
dynamical variable. Actually, as already discussed in
ref.~\cite{Mukohyama2000b}, if we fix the position of the brane by hand
as in the Gaussian normal coordinate system, then it is in general
inconsistent with convenient gauge choices in the bulk like a
generalized Regge-Wheeler gauge~\footnote{In the literature it is
sometimes called a generalized longitudinal gauge.}. In other words, as 
done in refs.~\cite{KIS,Mukohyama2000b}, we can construct $D$-gauge
invariant variables from the perturbation of the position of the brane,
and they are physical degrees of freedom independent of $D$-gauge
invariant variables in the bulk. The former gauge invariant variables
are concise in the sense that it is localized on the brane, and the
later variables can be expressed most concisely by the master variables
introduced in ref.~\cite{Mukohyama2000a}. Hence, the inclusion of the
brane position as a dynamical variable provides us with the most concise
configuration space.

Now let us illustrate the above arguments about $D$-gauge-invariant
variables by using some equations. For simplicity we consider
perturbations around a background with $3$-dimensional plane symmetry in
$5$-dimension. Namely, following the notation in
ref.~\cite{Mukohyama2001}, we  consider the metric 
%
\begin{equation}
 ds_5^2 = g_{MN}dx^Mdx^N = (g_{MN}^{(0)} + \delta g_{MN})dx^Mdx^N
\end{equation}
and the imbedding relation 
%
\begin{equation}
 x^M = Z^M(y) = Z^{(0)M}(y) + \delta Z^M(y),
\end{equation}
where the background is specified by a plane-symmetric background metric
%
\begin{equation}
 g^{(0)}_{MN}dx^Mdx^N = 
        \gamma_{ab}dx^adx^b + r^2\sum_{i=1}^3(dx^i)^2
\end{equation}
and such background imbedding functions $Z^{(0)M}(y)$ that $Z^{(0)a}$ 
depend only on $y^0$ and that $Z^{(0)i}=y^i$. Here, the two-dimensional
metric $\gamma_{ab}$ and the function $r^2$ are assumed to depend 
only on the two dimensional coordinates $\{x^a\}$. As for perturbations,
since in the linear order the perturbations of the position of the
hypersurface are decoupled from vector and tensor perturbations, we
consider scalar perturbations: 
%
\begin{eqnarray}
 \delta g_{MN}dx^Mdx^N & = & \int d^3{\bf k}\left[
        h_{ab}Ydx^adx^b 
        + 2h_{(L)a}V_{(L)i}dx^adx^i
        \right.\nonumber\\
 & &    \left.
        + (h_{(LL)}T_{(LL)ij}+h_{(Y)}T_{(Y)ij})dx^idx^j\right],
        \nonumber\\
 \delta Z_Mdx^M & = & \int d^3{\bf k}\left[
        z_aYdx^a
        + z_{(L)}V_{(L)i}dx^i\right],
\end{eqnarray}
where $Y=\exp(-i{\bf k}\cdot{\bf x})$, $V_{(L)i}=\partial_iY$,
$T_{(LL)ij}=2\partial_i\partial_jY+(2{\bf k}^2/3)\delta_{ij}Y$ and 
$T_{(Y)ij}=\delta_{ij}Y$, and all coefficients are supposed to depend
only on the $2$-dimensional coordinates $\{x^a\}$ of the orbit space. 
Here, ${\bf x}$ denotes coordinates $\{x^i\}$ of the three-dimensional
plane ($i=1,2,3$), and ${\bf k}$ represents the momentum $\{ k_i\}$
along the plane. Hereafter, we omit ${\bf k}$ in most cases. It is easy
to see how the coefficients $\{h's, z's\}$ transform under the $5$-gauge
transformation and to construct $5$-gauge-invariant
variables. Therefore, we obtain the following $5$-gauge-invariant
variables. 
%
\begin{equation}
 \phi_a = z_a + X_a, \label{eqn:phi_a}
\end{equation}
and
%
\begin{eqnarray}
 F_{ab} & = & h_{ab}-\nabla_aX_b-\nabla_bX_a,
        \nonumber\\
 F & = & h_{(Y)} -X^a\partial_br^2+\frac{2k^2}{n}h_{(LL)},
  \label{eqn:F_abF}
\end{eqnarray}
where $X_a=h_{(L)a}-r^2\partial_a(r^{-2}h_{(LL)})$ and $\nabla_a$
represents the covariant derivative compatible with the $2$-dimensional
metric $\gamma_{ab}$. The former variables (\ref{eqn:phi_a}) correspond
to perturbations of physical position of the hypersurface $\Sigma$ and
its normal component $\phi_an^{(0)a}$ appears in the
doubly-gauge-invariant junction condition, where $n^{(0)a}$ is the
background unit normal to the hypersurface. The latter (\ref{eqn:F_abF})
correspond to gravitational perturbations in the bulk and can be most
concisely expressed in terms of the master variable $\Phi$ as 
%
\begin{eqnarray}
 F_{ab} & = & \frac{1}{r}\left(\nabla_a\nabla_b\Phi 
        -\frac{2}{3}\nabla^2\Phi\gamma_{ab}
        + \frac{1}{3l^2}\Phi\gamma_{ab}\right),
        \nonumber\\
 F & = & \frac{r}{3}\left(\nabla^2\Phi-\frac{2}{l^2}\Phi\right). 
\end{eqnarray}
The perturbed Einstein equation in the bulk is reduced to the following
simple equation called master equation: 
%
\begin{equation}
 r^2\nabla^a\left[r^{-1}\nabla_a(r^{-1}\Phi)\right]
        - {\bf k}^2r^{-2}\Phi = 0.
\end{equation}

In the generalized Regge-Wheeler gauge where $h_{(L)a}=h_{(LL)}=0$, the
$5$-gauge-invariant variables are given by $\phi_a=z_a$, $F_{ab}=h_{ab}$
and $F=h_{(Y)}$. On the other hand, in the Gaussian normal gauge where 
$z_a=z_{(L)}=h_{(L)a}n^{(0)a}=h_{ab}n^{(0)b}=0$, these are given by
$\phi_a=X_a$ and (\ref{eqn:F_abF}). Note that in the Gaussian normal
gauge, $\phi_a$ is expressed in terms of metric perturbation. Therefore,
it is evident that $\phi_a$ cannot be set zero even in the Gaussian
normal gauge since $X_a\ne 0$ in general. Actually, requiring $\phi_a=0$
in the Gaussian normal gauge is equivalent to requiring $z_a=0$ in the
generalized Regge-Wheeler gauge, which is not possible in general.

Of course, it is always possible to take the Gaussian normal coordinate
system. In this coordinate system, as illustrated above, the
$5$-gauge-invariant variable $\phi_a$ is expressed in terms of metric
perturbations. Hence, as done in ref.~\cite{Garriga-Tanaka} for a static
background by a gauge-dependent method, we need to extract degrees of
freedom of $\phi_an^{(0)a}$ from the metric perturbations. Classically,
this procedure should not be difficult since we can use equations of
motion. However, quantum mechanically, we have to be careful when we use
the equations of motion to reduce the action.

The next task in the future is to obtain the second-order action for
perturbations by using the doubly covariant action obtained in this
paper. After that, we need to obtain the corresponding reduced action by
using a formalism to treat constrained systems, eg. Dirac's
method~\cite{Dirac} or Faddeev-Jackiw method~\cite{Faddeev-Jackiw}. As
shown in ref.~\cite{Mukohyama1997}, perturbative behavior of the
Wheeler-de Witt wave function can be investigated by the usual quantum
field theory in curved spacetime with the reduced action.

\begin{acknowledgments}

 The author would like to thank Werner Israel for his continuing 
 encouragement and helpful discussions. This work was done during the
 stay in Canadian Institute for Theoretical Astrophysics. The author
 would be grateful to Lev Kofman for his warm hospitality. This work was
 supported by the CITA National Fellowship and the NSERC operating
 research grant.

\end{acknowledgments}


\end{document}